\documentclass[twocolumn,pre,superscriptaddress,showpacs]{revtex4}  

\usepackage{graphicx}
\usepackage{amssymb}

\begin{document}

\title{Relationship between dynamical heterogeneities and stretched
exponential relaxation}

\author{S. I. Simdyankin} 

\affiliation{Department of Chemistry, University of Cambridge,
 Lensfield Road, Cambridge CB2 1EW, UK}

\affiliation{D\'epartement de physique and Centre de recherche en physique
et technologie des couches minces, Universit\'e de Montr\'eal,
C.P. 6128, succ. Centre-ville, Montr\'eal (Qu\'ebec) H3C 3J7, Canada}

\author{Normand Mousseau} 

\affiliation{D\'epartement de physique and Centre de recherche en physique
et technologie des couches minces, Universit\'e de Montr\'eal,
C.P. 6128, succ. Centre-ville, Montr\'eal (Qu\'ebec) H3C 3J7, Canada}

\date{\today}

\begin{abstract}
We identify the dynamical heterogeneities as an essential prerequisite
for stretched exponential relaxation in dynamically frustrated
systems. This heterogeneity takes the form of ordered domains of
finite but diverging lifetime for particles in atomic or molecular
systems, or spin states in magnetic materials. At the onset of the
dynamical heterogeneity, the distribution of time intervals spent in
such domains or traps becomes stretched exponential at long time.  We
rigorously show that once this is the case, the autocorrelation
function of the renewal process formed by these time intervals is also
stretched exponential at long time.
\end{abstract}

\pacs{61.20.Lc, % Time-dependent properties;relaxation (for glass 
	        % transitions, see 64.70.Pf
      05.45.Xt, % Synchronization; coupled oscillators
      61.43.Fs, % Glasses
      05.45.Ra  % Coupled map lattices
     }

\maketitle

%-------------------------------------------------------------------------
%                       Main text
%-------------------------------------------------------------------------
%
%

\section{Introduction}

The stretched exponential relaxation (SER), $I(t) \sim
\exp[-(t/\tau)^\beta]$, describes the relaxation of pure glasses very
accurately, often over very wide ranges \cite{Phillips_96}. For
example, SER describes equally well stress experiments on amorphous
Se, centered on $10^3$ s, as well as spin-polarized neutron scattering
experiments, centered on $10^{-9}$ s, both with the same value of
$\beta$, over a time range of more than $10^{12}$.  SER also describes
accurately stress relaxation of the superconducting transition
temperatures in the cuprates, probably associated with the two-gap
domain structure revealed by high resolution scanning tunneling
microscopy \cite{Phillips_2003}; moreover, the measured value of
$\beta$ is the predicted value, 3/5, for YBCO, by far the most
extensively studied cuprate, with the highest quality samples. The
agreement is such that it has been recently proposed that the SER be
used as an independent measure of sample quality in these materials
\cite{Phillips_2000}. Phillips' review \cite{Phillips_96} gives many
more examples and demonstrates the power of SER as a guide to
understanding the dynamical properties of intrinsic glasses,
especially network glasses.  One of the most spectacular examples was
the successful prediction \cite{Phillips_97} of the long- and
short-preparation-time values of $\beta$ of orthoterphenyl (OTP), as
measured by multidimensional NMR.  OTP is the purest organic glass
former available commercially; the predicted values were 0.43 and
0.60, and the observed values were 0.42 and 0.59.

It appears that SER may arise as a result of spatial inhomogeneities
that are quenched into macroscopically pure, non-phase separated
glasses on a microscopic scale as traps or sinks that appear as part
of the glass-forming process.  The existence of these inhomogeneities is difficult to detect experimentally, but they readily account for SER,
and predict correctly the values of beta in many experiments
\cite{Phillips_96}.  Dynamical heterogeneities have been seen
experimentally and numerically in glasses and supercooled
liquids~\cite{Donati_99,Debenedetti_01}, spin-glass
models~\cite{Mansfield_2002,Garrahan_2002}, Lennard-Jones alloy
mixtures with ellipsoidal probes \cite{Bhattacharyya_2002} as well as
in coupled-chaotic systems such as diode-resonator arrays
~\cite{Hunt_2002}.  In particular, the concept of dynamical
heterogeneities has emerged as critical for the description of the
microscopic dynamics associated with the dramatic slowing down of the
relaxation and diffusion processes as the temperature approaches
$T_g$, the glass-transition temperature
\cite{Richert_2002,Sillescu_99,cristanti02}.

These spatial heterogeneities occur over time scales that are
long with respect to the basic unit time of the systems --- one step
in coupled maps and a lattice vibration in supercooled liquids ---
but shorter than the macroscopic relaxation, where materials are
known to be homogeneous. As the temperature or the driving force is
decreased, these regions become more and more stable and might
overtake the whole sample at a transition temperature. 
Although the nature of spatial heterogeneities and their relation to
the macroscopic dynamics remains incompletely understood, there is a
considerable amount of work that relates directly to this issue. The
mode-coupling theory describes the dynamics from the liquid phase into
the supercooled regime and makes quantitative predictions that have
been verified numerically in many systems~\cite{chong01,chong02}. It
cannot yet describe accurately the very long time behavior in the
supercooled regime, however~\cite{chong01}.
Numerically, dynamical heterogeneities have been studied using a wide
range of criteria tied various aspects of this
feature\cite{Donati_99,VollmayrLee_2002,Caprion_2000}. 
Recently, however, some groups have focused on the renewal or mean
first-passage time (MFPT)~\cite{Allegrini_1999} and the waiting-time
distribution (WTD) which provide some link to the structure of the
energy landscape~\cite{Doliwa_2003,Denny_2003}.

Although both the dynamical heterogeneities and the stretched
exponential behavior are characteristics of many frustrated systems
(e.g. supercooled liquids \cite{Xia_2001}), it is not clear what the
relation between these two properties is.  In this paper, we address
this question and show that dynamical heterogeneities can be directly
responsible for the observation of stretched exponential relaxation in
coupled map lattices. We also show that these results are applicable
to supercooled liquids and can provide a microscopic basis for the
stretched exponential relaxation in these systems, in line with the
observations of Denny {\it et al.} on the long time dynamics of
meta-basin hopping~\cite{Denny_2003}.

Recently, Hunt {\it et al.} found that the distribution of the
renewal time of a certain stochastic process measured in a
one-dimensional lattice of coupled diode resonators could be fitted to
a stretched exponential function over 6 orders of magnitude
\cite{Hunt_2002}. 
Simulations on a related model of diffusively coupled via short range
interactions nonlinear maps show that the fit can be extended to more
than 9 orders of magnitude, ruling out any other power law or simple
combination of pure exponentials
\cite{Hunt_2002,Simdyankin_2002_PRE}. The quality of the fit in this
system is sufficient to distinguish the region of the parameter space
where the dynamics is described by power-law distribution coupled with
an exponential cut-off from the really stretched exponential
distributions. (Ref. \cite{Simdyankin_2002_PRE} contains background
information important for understanding the results of the present
work and will be referred to as [I] in the following.)

These simulations [I] also demonstrated the importance of the system's
size on its dynamics. While stretched exponentials are sufficiently
robust and are observed even in some small samples, the quality of
such fits is much poorer than in the big samples. This is in direct
analogy with real experimental results and supports the observation
that the quality of the samples is crucial for obtaining unambiguous
stretched exponential behavior and meaningful values of the exponent
$\beta$ \cite{Phillips_96,Phillips_2000}.

Using simulations on this and other models, as well as some analytical
results, we show here that (1) the stochastic process studied in
Refs.~\cite{Hunt_2002} and [I] can be identified with a renewal
process~\cite{Godreche_2001}, (2) a stretched exponential distribution
of the renewal time implies a similar shape for the decay of the
two-point auto-correlation function, and (3) renewal processes are
present in a supercooled liquid and play a central role in its
dynamics.

Granted the universal features of SER, readers unfamiliar with chaos
theory and coupled maps may not see immediately the relevance of these
mathematical tools to understanding the dynamical properties of
supercooled liquids and glasses.  Even the successful predictions of
the trap (sink) model \cite{Phillips_96,Hughes_RW} appear to be
unexpected.  For instance, the survival probability of a random walker
in the presence of randomly distributed static traps crosses over from
exponential to SER only at longer times where very few ($\sim
10^{-30}$ for 10\% traps) of the walkers have survived
\cite{Barkema_2001}.  In our examples of steady-state dynamics in
coupled maps, the analogous crossover in trap-time distributions
$\rho(t)$ occurs near $10^{-1}$ to 10$^{-2}$ $\times \max\{\rho(t)\}$
depending on the value of a control parameter.  Even the latter number
is small compared to the crossovers observed, for example, for
relaxation of the first peak in the structure factor in molecular
dynamics simulations (MDS) of crystallization-avoiding soft sphere
mixtures \cite{Roux_1989}, which occur near 3/4 of the maximum
value. The reason for these differences is that the systems chosen for
MDS already represent prepared states with strong glass-forming
tendencies.  The random walker with random traps model contains
spatial inhomogeneities, which we will show represent an essential
feature necessary for SER, and which are absent from some popular
models of static glass structure such as mode-coupling theory
\cite{Gotze_92}.  That model does not contain the collapse of phase
space that occurs as the liquid is supercooled to the glass
transition, and that is why SER appears only at very long times.
However, it does contain the correct dimensional dependence of
diffusive behaviour that leads to the relation $\beta = d/(d + 2)$,
which is important for comparisons to experiment. 
That the one-dimensional coupled map lattice discussed here reaches a steady-state much more rapidly than the random walker is a clear indication that this model is physically relevant. The way in which this happens was illustrated
in Fig. 8 of [I]: the surviving periodic domains are those that have
avoided chaotic regions.

We also emphasize that in this simple model, the behavior of the
stretching exponent $\beta$ as a function of a control parameter $r$
is similar to the dependence of $\beta$ on the temperature in both
3D \cite{Ogielski_1985} and 1D \cite{Mansfield_2002} spin-glass
models.  This similarity not only captures the qualitative tendency 
for $\beta$ to decrease with decreasing $r$ (or $T$), but also the 
quantitative agreement in the low-$r$ (or $T$) limit $\beta \approx 1/3$.
The same lowest value of $\beta$ was obtained in studies of random 
walks on fractals \cite{Campbell_85, Almeida_2000, Jund_2001}.

\section{Results}

\subsection{Properties of the coupled map model}

\begin{figure}
\centerline{\includegraphics[width=7.5cm]{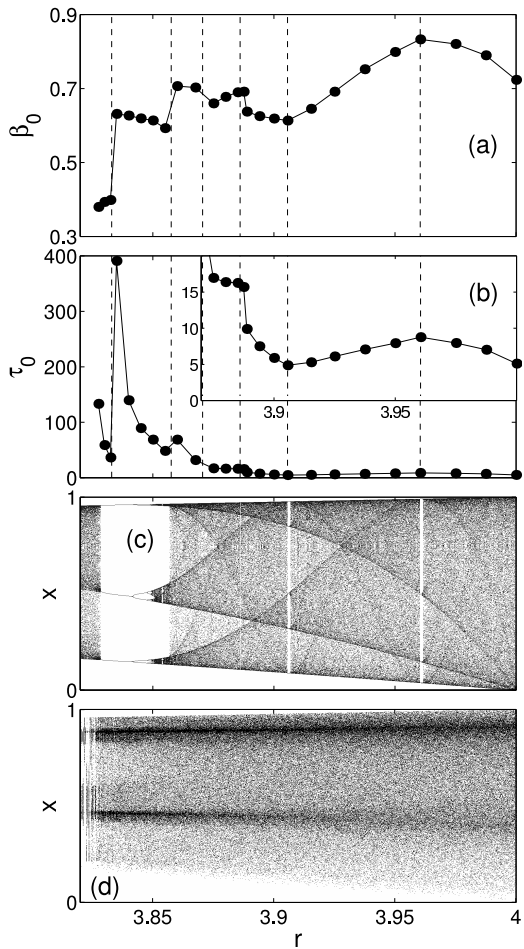}}
\caption{$\beta_0$ (a) and $\tau_0$ (b) as functions of the logistic
map parameter $r$. Dots correspond to the least squares-fit values and
the thin solid line is drawn as a guide to the eye. The vertical
dashed lines demarcate the discontinuities in $\beta_0(r)$ and
$\tau_0(r)$. They correspond very well to the positions of the
periodic cascades in the bifurcation diagram of the uncoupled logistic
map shown in (c).  (d) shows the bifurcation diagram for a site in the
coupled map lattice.  Inset in (b) magnifies the right hand side of
the main plot where the variation of $\tau_0$ with $r$ is not
distinguishable.  The accuracy of $\beta_0$ and $\tau_0$ obtained by a
least-squares fit to the long time part of the trap-time distribution
is within 15\%.}
\label{fig:betavsr}
\end{figure}

The nonlinear model of Ref.~\cite{Hunt_2002}, used to reproduce
qualitatively the experimental system, is a one-dimensional chain of
$N$ diffusively coupled nonlinear deterministic maps, $f(x)$, with a
coupling constant $\alpha$ and periodic boundary conditions.
The time evolution of this system is discrete and is described by the
following iterative equation,
\begin{equation}
x_n(t+1) = (1-\alpha) f(x_n(t)) + 
                  \frac{\alpha}{2} \left\{ f(x_{n-1}(t)) +
                                           f(x_{n+1}(t)) \right\}.
\label{eq:cml} 
\end{equation} 
which, given initial conditions $x_n(0)$ for each site
$n=1,2,\dots,N$, generates a time series $\{x_n(t)\}$.
The interaction is totalistic and involves only the nearest neighbors.
Following [I], in this study we use the logistic map $f(x)=rx(1-x)$.

Following the analysis of the experimental data, the stochastic process
of interest is represented by a coarse-grained variable defined as
\begin{equation}
\sigma_n(t) = \mathrm{sgn}[x_n(t)-x_\mathrm{thr}], \; t = 0,2,4,\dots
\label{eq:renew}
\end{equation}
where the quantity $x_n(t)$ is defined in Eq.~(\ref{eq:cml}) and
$x_\mathrm{thr}$ is a certain threshold value. 
The basic dynamics of the coupled oscillators being period two, the
analysis is also done over even (or odd) time steps.

This model shows a stretched exponential distribution of the renewal
(or trap) times, which are defined in the next section
\ref{sec:renewals}, over a wide range of the control parameter $r$,
which plays a role akin to temperature as can be seen in
Fig.~\ref{fig:betavsr}.  This figure shows the stretching exponent
$\beta_0$ and the time scale $\tau_0$ as a function of the control
parameter $r$.  The values of $\beta_0$ and $\tau_0$ were obtained by
least-square fitting the long-time part of the trap-time distributions
for a chain of $N=10,000$ coupled logistic maps $f(x)=rx(1-x)$ with
the expression $A\exp[-(t/\tau_0)^{\beta_0}]$.
Although $\beta_0(r)$ and $\tau_0(r)$ are non-monotonic, the overall
behavior is similar to the results in both 3D \cite{Ogielski_1985} and
1D \cite{Mansfield_2002} spin-glass models.
This broad similarity not only captures the qualitative tendency for
$\beta$ to decrease with decreasing $r$ (or $T$), but also the
quantitative agreement in the low-$r$ (or $T$) limit $\beta \approx
1/3$; the same limit as that obtained in studies of random
walks on fractals \cite{Campbell_85, Almeida_2000, Jund_2001}.

Although the bifurcation diagram for a site in the coupled map lattice
(Fig.~\ref{fig:betavsr}(d)) appears to be uniform with respect to $r$
down to $r\approx3.8275$, the non-monotonicity of $\beta_0(r)$ and
$\tau_0(r)$ most probably arises from the fact that for the isolated
logistic map and the values of $r<4$ chaotic orbits are interspersed
with period cascades seen in Fig.~\ref{fig:betavsr}(c), i.e. the
approach to full chaos as $r$ goes to 4 is not uniform. (For an
explanation of the concept of bifurcation diagrams see
e.g. Ref.~\cite{Ott_CDS}.)
Interestingly, the lowest value of $\beta_0$ is attained within the
widest window of the period-three cascade in the bifurcation diagram
of the uncoupled logistic map.
About and below $r=3.82$, in the coupled map chain, very narrow
chaotic windows are also interspersed with periodic ones and the
stretched exponential behavior of the trap-time distribution, with
$\beta_0$ greater or about 1/3, is recovered only for some values of
$r$.
This interesting behavior suggests that SER is uncovering a new aspect
of the dynamics of coupled map lattices. A full analysis of this new
behavior lies outside the framework of this paper, but we plan to
return to studying it elsewhere.

In other models and experiments, the time scale $\tau$ monotonically 
grows with decreasing temperature. 
It is not the case here, although the values of $\tau_0$ are greatest
at the low-$r$ (and low-$\beta_0$) end, following the general trend
observed elsewhere.

\subsection{Renewal processes in coupled maps}
\label{sec:renewals}

The stochastic process $\sigma_n(t)$ is a renewal
process~\cite{Godreche_2001} provided that the time intervals (renewal
times) between two subsequent zero crossings are statistically
independent random variables.
We tested the independence of these time intervals for the present
model, Eq.~(\ref{eq:renew}), by calculating the distributions of the
time intervals following time intervals of a specified length.
The distributions were identical regardless of the length of the 
preceding time intervals.

In general, a renewal process is defined so that $\sigma(t)=+1$ for
$t_0 < t \le t_1$, $\sigma(t)=-1$ for $t_1 < t \le t_2$,
$\sigma(t)=+1$ for $t_2 < t \le t_3$, and so on.
The renewal processes are the simplest possible stochastic processes
readily susceptible of probability theoretical analysis
\cite{Feller_IPTA}.
The dynamics represented by such processes can be understood in terms
of transition probabilities between the two possible states
$\sigma=\pm1$ \cite{Gielis_95}.

\begin{figure} 
\centerline{\includegraphics[width=7.5cm]{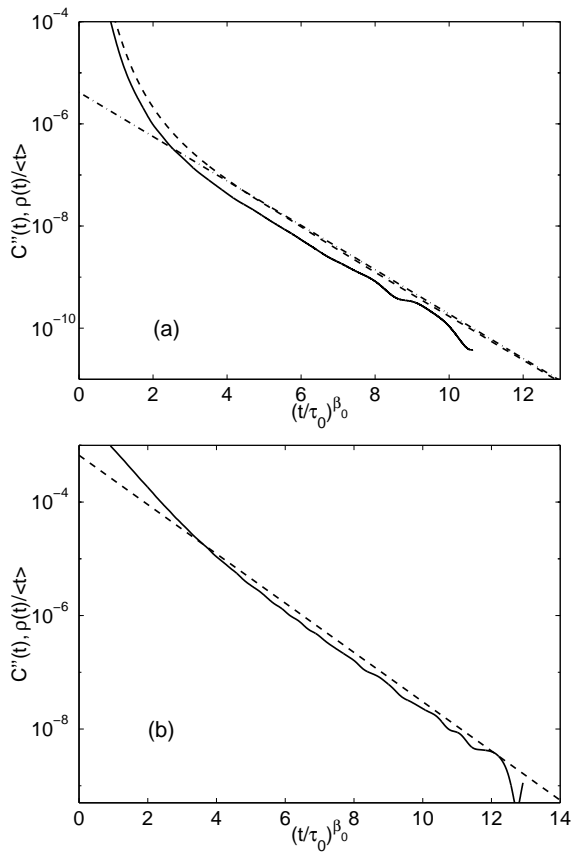}}
\caption{(a) Second derivative $C''(t)$ of the autocorrelation
function $C(t)$ of the coarse-grained orbit $\sigma(t)$,
Eq.~(\ref{eq:renew}) with $r=3.83$, (solid line) compared to the
asymptotic behavior $\rho(t)/\langle t\rangle$ given by
Eq. ~(\ref{eq:ddc}) (dashed line). Dash-dotted lines shows the
stretched exponenetial fit to the long-time part pf $\rho(t)$ with
$\tau_0 = 59\pm9$ and $\beta_0 = 0.39\pm 0.05$).  (b) The same as in
(a), but for the renewal process computer generated with the renewal
times distributed according to Eq.~(\ref{eq:strexp}) with $\tau_0 = 5$
and $\beta_0 = 0.4$.  In this case, the lines corresponding to the
dashed and dash-dotted lines in (a) coincide.}
\label{fig:d2c}
\end{figure}

In [I], it is shown that the existence of very long renewal time
intervals $t_{n+1}-t_n$ associated with the stretched exponential
distribution can be attributed to the presence of time-limited ordered
domains (or traps) with dominant spatial period of 4 sites.
Fig.~\ref{fig:d2c}(a) shows a typical distribution of the renewal
time intervals (or trap times) for this model.

\begin{figure} 
\centerline{\includegraphics[width=7.5cm]{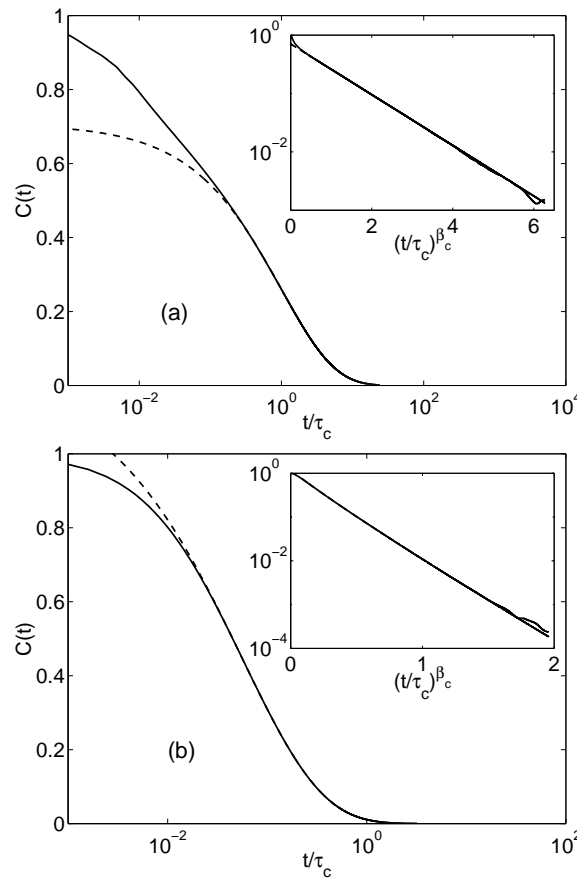}}
\caption{(a) Autocorrelation function $C(t)$ of the coarse-grained
orbit $\sigma(t)$, Eq.~(\ref{eq:renew}) wiht $r=3.83$, (solid line)
and the stretched exponenetial fit to the long-time part of $C(t)$
with $\tau_c \approx 1250$ and $\beta_c \approx 0.58$ (dashed
line). The inset shows the same curves, but with the logarithmic
vertical axis.  (b) The same as in (a), but for the renewal process
computer generated with the renewal times distributed according to
Eq.~(\ref{eq:strexp}) with $\tau_0 = 5$ and $\beta_0 = 0.4$.  In this
case the stretched exponential fit to $C(t)$ is with $\tau_c \approx
69$ and $\beta_c \approx 0.54$.}
\label{fig:corfun}
\end{figure}

These results suggest that the dynamical heterogeneity manifested by
the presence of time-limited ordered domains could be at the origin of
the stretched exponential relaxation in supercooled liquids and
glasses and the renewal processes could provide a simple picture for
describing heterogeneous dynamics in these condensed phases.
In order to support these two affirmations, we first need to show that
(1) there is a one to one correspondence between renewal processes and
their time auto-correlation functions, because experimentally
measurable relaxation responses are mathematically described by
auto-correlation functions of certain dynamical variables and not
directly by non-observable renewal processes, and that (2) the renewal
processes are present in supercooled liquids and glasses and can be
calculated effectively, exhibiting unambiguous stretched exponential
distributions of renewal time intervals.

The relation between the auto-correlation function 
\begin{equation}
C(t) \equiv <\sigma(t')\sigma(t'+t)>_{t'}
\label{eq:corfun}
\end{equation}
of a renewal process $\sigma(t)$ and the distribution $\rho(t)$ of
time intervals $t_{n+1}-t_{n}$ between the zero crossings (or, in other
words, renewals) of this process was recently studied for a range of
distributions~\cite{Godreche_2001}.
Following this approach, Godr\`eche and Luck also
showed~\cite{godreche2002} that $C(t)$ corresponding to the stretched
exponential distribution
\begin{equation}
\rho(t) = \frac{\beta_0}{\tau_0 \Gamma(\beta_0^{-1})}
\exp[-(t/\tau_0)^{\beta_0}]
\label{eq:strexp}
\end{equation}
can also be represented by a stretched exponential at long time 
\begin{equation}
C(t) \approx C_{\mathrm{s.e.}}(t) \equiv A\exp[-(t/\tau)^{\beta}], \; t
\to \infty 
\label{eq:strexpC}
\end{equation}

We can Laplace transform the distribution of intervals $\rho(t)$,
obtaining a function with a small singular part, $\hat{\rho}(u) = 1 -
\langle t \rangle u + \cdots + \hat{\rho}_{\mathrm{sing}}(u)$, $u \to
0$~\cite{footnote1}, where $\langle t \rangle = \tau_0
\Gamma(2/\beta_0)/\Gamma(1/\beta_0)$, is the first moment of
$\rho(t)$.
Also from Ref.~\cite{Godreche_2001}, we know that the Laplace transform of
the auto-correlation function is given by 
\begin{equation}
  \hat{C}_{\mathrm{eq}}(u) = \frac{1}{u} \left( 1 -
  \frac{2(1-\hat{\rho}(u))}{\langle t \rangle u (1+\hat{\rho}(u))}\right)
\end{equation}
Using  the expression above for $\hat{\rho}(u)$, one finds, for the
singular part of the correlation function at equilibrium
\begin{equation}
 \hat{C}_{\mathrm{eq,sing}}(u) \approx
 \frac{\hat{\rho}_{\mathrm{sing}}(u)}{\langle t \rangle u^2}
\end{equation}
Since the behavior at large time is governed by its singular part,
we get
\begin{equation}
C''_{\mathrm{eq}}(t) \approx \rho(t)/ \langle t \rangle.
\label{eq:ddc}
\end{equation}
This implies that for a stretched exponential $\rho(t)$ in the
asymptotic limit $C_{\mathrm{eq}}(t) \propto \rho(t)$\cite{godreche2002}.

The range of relevance for this analytical result can be confirmed
numerically by studying the statistical properties of a
computer-generated sequence of (pseudo)random numbers distribution
according to the stretched exponential probability density function
(pdf) Eq.~(\ref{eq:strexp}).
Using the fundamental transformation law of probabilities
\cite{Devroye_NURVG}, given a computer-generated sequence of
(pseudo)random numbers $\{\theta\}$ distributed uniformly in the
interval $0 \le \theta \le 1$, we apply a transformation rule $t =
\tau_0 [Q^{-1}(1/\beta_0,\theta)]^{1/\beta_0}$ so that the resulting
sequence $\{t\}$ is distributed according to the stretched exponential
pdf $\rho(t)$.
In order to compute the inverse regularized incomplete gamma function
$Q^{-1}(a,\theta)$, we used the software from
Refs.~\cite{DiDonato_87,DiDonato_86}.

Given a sequence of time intervals distributed according to a
prescribed pdf, it is straightforward to construct the corresponding
renewal process and its autocorrelation function and second derivative.
The second derivative is calculated numerically, after applying a
smoothing function before each derivative to improve the quality of
the operation.

The resulting $C''(t)$ and $C(t)$, shown in Figs.~\ref{fig:d2c} and
\ref{fig:corfun} respectively, demonstrate clearly that the
correspondence between $\rho(t)$ and $C''(t)$ holds for long times for
$C(t)$ calculated both for the coupled map lattice model,
Eq.~(\ref{eq:cml}) with $f(x)=rx(1-x)$, $r=3.83$, $\alpha=0.25$ and
$N=1000$, and for a renewal process where the renewal time-interval
distribution is stretched exponential by prescription.
$C''(t)$ obtained numerically agrees with $\rho(t)/\langle t \rangle$,
Eq.~(\ref{eq:ddc}), without any free parameter within a multiplicative
factor of the order of unity.

Interestingly, for the range measured, $C(t)$ also follows the stretched
exponential form, albeit with somewhat different parameters $\tau_c$
and $\beta_c$.
This indicates that the true asymptotic regime for $C(t)$ or $\rho(t)$
may be reached at longer time with a gradual approach towards this
long time behavior.
This agrees with the theories (see e.g. \cite{Hughes_RW}) where the
stretched exponential functions are accompanied by time-dependent
multiplicative factors.
For practical purposes, however, it is remarkable that the stretched
exponential fit to the correlation function agrees with $C(t)$
starting with a well-observed experimentally value between 3/4 and
1/2~$\times \max\{C(t)\}$ and follows with very high accuracy for
several orders of magnitude to the regime beyond the reach of the
experiments.

\subsection{Molecular dynamics}

Having established the correspondence between the distribution of time
intervals of renewal processes and their auto-correlation functions,
we now show that a renewal process can also be identified in a
supercooled liquid and it appears to provide a simple and elegant
description for the stretched exponential relaxation.
In order to do so, we have performed molecular dynamics simulations of
a 16000-particle single-component system where the interactions
between the particles are described by a pair potential labeled Z2 in
Ref.~\cite{Doye_2002_Z1and2}.
This potential is similar to that introduced initially by
Dzugutov~\cite{Dzugutov_92} but provides a longer-lived metastable
supercooled state prior to crystallization.
Otherwise the properties of the Z2 liquid are similar to those of the
Dzugutov liquid, in particular it exhibits dynamical heterogeneities
formed by short-lived clusters composed of connected icosahedra
\cite{Dzugutov_2001}.
The simulations are performed at $T=0.68$, below the melting
temperature of $T_m=0.70 \pm 0.05$ with a time step of 0.01 in reduced
Lennard-Jones units \cite{footnote2}. We take measurement over the
16~000 atoms with the number density $\rho = 0.85$ for 7 million time
steps, stopping before the system starts crystallizing. This
simulation time would correspond to a 150-ns simulation for argon.

A renewal process representing the atomic dynamics in the liquid can
be constructed so that each renewal time interval correspond to the
time a particle spends in a dynamical trap. This is identical to the
first passage time, in the parlance of Allegrini {\it et
al.}~\cite{Allegrini_1999}, but differs formally from the waiting time
of Doliwa and Heuer~\cite{Doliwa_2003} and Denny {\it et
al.}~\cite{Denny_2003} which focuses on hops between energy basins.
We say that particle $i$ is trapped as long as its displacement
\begin{equation}
d_i(t) = \left|\mathbf{r}_i(t)-\mathbf{r}_i(0)\right|
\label{eq:renewmd}
\end{equation}
measured starting from an arbitrary initial position $\mathbf{r}_i(0)$
remains less than a certain threshold value $d_\mathrm{thr}$.
The moment $t=t_\mathrm{thr}$ of surpassing the threshold is counted
as a renewal and the value of $\mathbf{r}_i(0)$ is reset to
$\mathbf{r}_i(t_\mathrm{thr})$ as well as the time origin for the next
trap.

\begin{figure}
\centerline{\includegraphics[width=7.5cm]{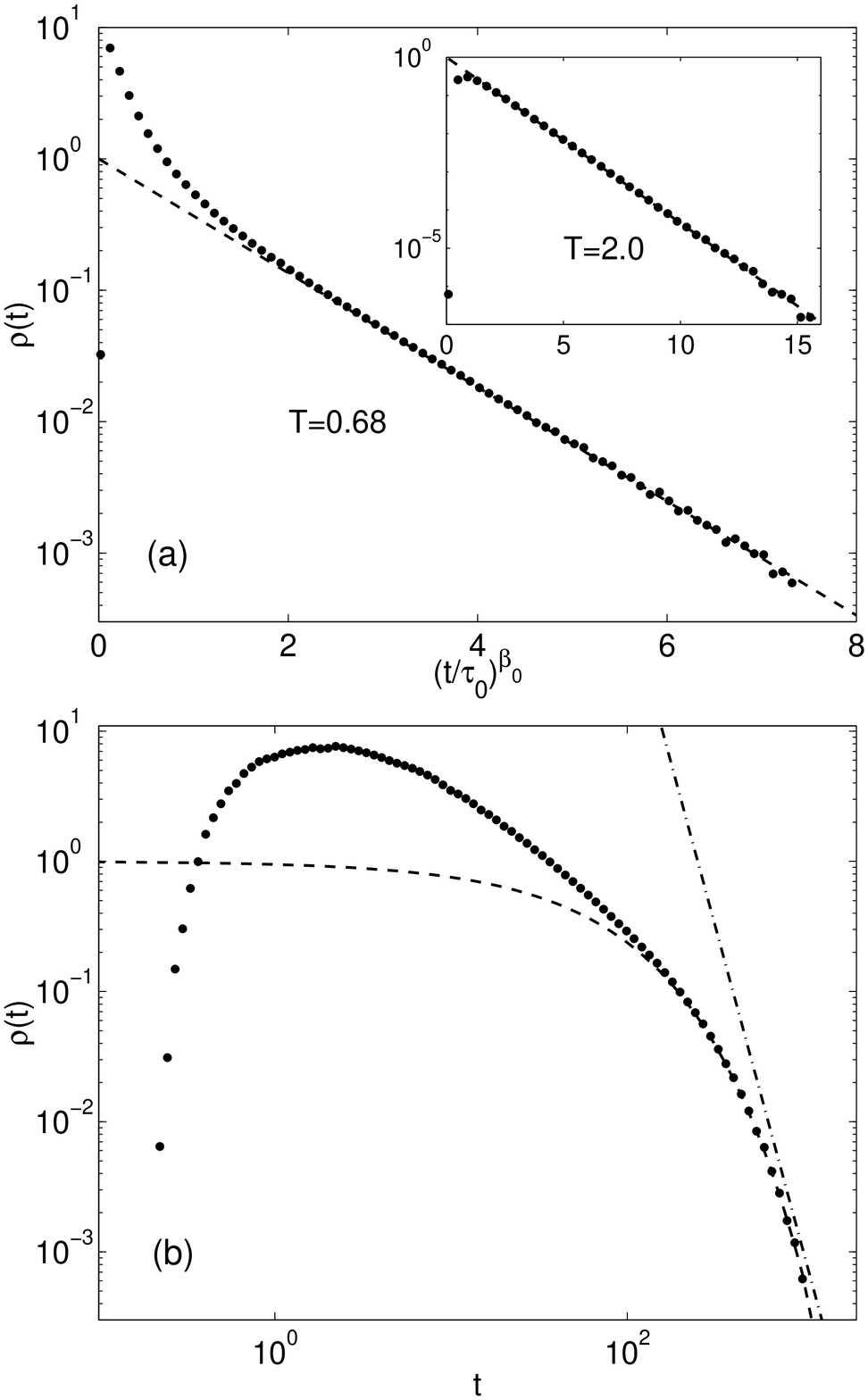}}
\caption{(a) Dots: distributions of trap-time intervals for low
(T=0.68, each data point corresponds to an average over 500 
time steps) and high temperatures (inset, T=2). Dashed lines:
corresponding stretched exponential fits with $\tau_0=60\pm5$,
$\beta_0=0.71\pm0.05$ and (inset) $\tau_0=1.22\pm0.05$,
$\beta_0=1\pm0.002$. (b) Dots: the same as in (a) for T=0.68 presented
on a log-log scale over a mesh with logarithmic time intervals.
Dashed line: the same as in (a) for T=0.68. Dash-dotted line: power
law $\propto 1/t^5$.}
\label{fig:histhops}
\end{figure}

The corresponding trap-time distribution (see Fig.~\ref{fig:histhops})
is not particularly sensitive to the choice of $d_\mathrm{thr}$ as long
as it is of the order of the effective atomic diameter.
For definitiveness, we choose $d_\mathrm{thr} = 2\pi/Q_0 \approx
0.88$~\cite{footnote2}, where $Q_0$ is the position of the main peak
in the static structure factor $S(Q)$ and neglect the temperature
dependence of this length scale. A threshold of $\approx 0.88$ is
about 4 times larger than the critical value estimated by Allegrini
{\it et al.} in a binary Lennard-Jones system. While the focus of
Ref.~\cite{Allegrini_1999} is on the inertial regime, which takes
place as small distance, on the order of 0.1 atomic radius, we are
interested in the atomic displacement leading to a change in the
configuration.
Figure ~\ref{fig:histhops}(a) also shows in inset that the trap-time
distribution is exponential at the temperature well above the melting
point ($T_m = 0.70 \pm 0.05$) and its long-time tail well agrees with
a stretched exponential fit with $\beta_0 \approx 0.71$ in the
supercooled regime.
In order to be considered as a fully Markovian process, we also assess
the correlation between subsequent trapping time intervals, as was
done for the coupled map lattice. Comparing trap distributions
considering only the subset of events following traps of various
length, again, we find no correlation between renewal time intervals.
The value of $\beta_0$ agrees well with $\beta = 0.70 \pm 0.05$ for
the self part of the intermediate scattering function $F_s(Q_0,t)$
calculated at the same temperature and for the same length scale.
(For the definition of $F_s(Q,t)$ see, e.g., Ref. \cite{Hansen_TSL}.)
$F_s(Q_0,t)$ for the Z2 liquid behaves similarly to the same quantity
calculated for the supercooled Dzugutov liquid \cite{Dzugutov_94} and
is not shown here.

These results appear at odds with the analysis of Allegrini {\it et
al.} who report a power-law distribution for the long-time tail of the
renewal time~\cite{Allegrini_1999}, although admitting that the
asymptotic cutoff in the distributions is difficult to resolve
numerically.  Fig. ~\ref{fig:histhops}(b) shows our simulation results
at $T=0.68$ plotted on a log-log scale with a tentative fit of the
long-time behavior. It is clear that the stretched exponential
provides a much better fit to the simulation data. This difference
between the two simulations can be due to both a physical range of
parameters and the precision of the simulation. First, Allegrini and
collaborators focus on the inertial length-scale while the results
presented here require atomic diffusion, not only thermal vibrations;
second, focusing on a single temperature, our simulation uses a 
larger cell and averages over a time scale roughly 6 times longer.

Denny {\it et al.}~\cite{Denny_2003} also note that the distribution 
in Ref.~\cite{Allegrini_1999} may not be a power law. 
Although the waiting time distribution in Ref.~\cite{Denny_2003} differs
formally from that is studied here, at long time they can be comparable
since the escape of a particle from a sphere of radius of the order of
the atomic diameter is likely to correspond to the transition of the
phase point in the configuration space between energy metabasins.
We note however, that the log-normal form of the waiting time
distribution proposed by Denny {\it et al.} results in a correlation
function that can only be \textit{approximated} with SER, whereas the
stretched exponential first passage-time distribution \textit{results}
in a correlation function that is also stretched exponential at long
time.

\section{Conclusion and discussion}

As was shown in [I], traps in coupled map lattices correspond to
ordered short-lived domains.
In direct analogy, it is natural to expect that the trap condition
$d_i(t) < d_\mathrm{thr}$, for $d_i(t)$ from Eq.~(\ref{eq:renewmd}),
is satisfied as long as particle $i$ is literally trapped within a
relatively stable environment - a ``cage''.
Thus formed renewal process and the associated trap-time distribution
and the autocorrelation function provide a simpler description of the
structural relaxation in a condensed matter system than, e.g., the
cage correlation function \cite{Rabani_99} recently used to
demonstrate the stretched exponential relaxation in a Lennard-Jones
massively defective crystal.

In particular, traps in the coupled map lattice show only very short
spatial extension that does not seem to diverge as the overall
dynamics of the system slows down. As $\beta_{0}$ goes from $0.6>\beta_0>0.5$
to $0.4>\beta_0>0.3$, the longer traps go from about 120 to more than 30,000
time steps but the characteristic time-averaged width of these traps
only quadruple, from 2 to 8 sites [I].
These observations are consistent with the behavior of dynamical
heterogeneities observed in a number of model systems
\cite{Mansfield_2002,Garrahan_2002,VollmayrLee_2002,Dzugutov_2001}.

We note that the only model that rigorously derives the stretched
exponential long time asymptotic behavior is that of a random walker
in a system with static traps (see e.g. Ref.~\cite{Phillips_96}). 
Ref.~\cite{Hughes_RW} cites both upper \cite{Kayser_1983} and lower
\cite{Grassberger_82} bounds for the asymptotic behavior of the
survival probability.
Both bounds have the same $\beta$, but different $\tau$.
It is not obvious, however, that the above theory directly applies to
the results of this works.
Nevertheless, our results are obtained in terms of first passage
statistical distributions of the type encountered in the static-trap
model and the use of similar mathematical techniques may shed more
light on the nature of the dynamics in the models presented here.

In conclusion, we focused on a relation between dynamical
heterogeneities and stretched exponential relaxation in two models: a
coupled chaotic oscillator system and a supercooled liquid.  Without
insisting that the former model represents the actual behavior of real
liquids, we have shown that the asymptotic long time behavior of the
autocorrelation function of a renewal process is stretched exponential
provided that the distribution of the renewal time intervals $\rho(t)$
is also stretched exponential at long time, i.e. $\rho(t) \approx
\exp[-(t/\tau_0)^{\beta_0}]$.
Using this relation, we have demonstrated that, at least in the two
model systems considered here, the relaxation dynamics can be
described in terms of well-defined dynamical traps, providing useful
insight regarding the absence of a diverging static length scale 
at the glass transition.

%\vspace{-0.5cm}

\section*{Acknowledgments} 

We thank C. Godr\`eche for his essential help in relating formally the
auto-correlation function to the distribution of renewal times and
J.~C.~Phillips for discussions, correspondence and valuable comments.
We acknowledge partial support from the Natural Sciences and
Engineering Council of Canada (NSERC) as well as the {\it Fonds Nature
et Technologie du Qu\'ebec}. These calculations were performed mostly
on the computer of the R\'eseau qu\'ebecois de calcul de haute
performance (RQCHP).  SIS is grateful to EPSRC for support. NM is a
Cottrell Scholar of the Research Corporation.

\bibliographystyle{apsrev}
%\bibliography{archive}

\begin{thebibliography}{48}
\expandafter\ifx\csname natexlab\endcsname\relax\def\natexlab#1{#1}\fi
\expandafter\ifx\csname bibnamefont\endcsname\relax
  \def\bibnamefont#1{#1}\fi
\expandafter\ifx\csname bibfnamefont\endcsname\relax
  \def\bibfnamefont#1{#1}\fi
\expandafter\ifx\csname citenamefont\endcsname\relax
  \def\citenamefont#1{#1}\fi
\expandafter\ifx\csname url\endcsname\relax
  \def\url#1{\texttt{#1}}\fi
\expandafter\ifx\csname urlprefix\endcsname\relax\def\urlprefix{URL }\fi
\providecommand{\bibinfo}[2]{#2}
\providecommand{\eprint}[2][]{\url{#2}}

\bibitem[{\citenamefont{Phillips}(1996)}]{Phillips_96}
\bibinfo{author}{\bibfnamefont{J.~C.} \bibnamefont{Phillips}},
  \bibinfo{journal}{Rep. Prog. Phys.} \textbf{\bibinfo{volume}{59}},
  \bibinfo{pages}{1133} (\bibinfo{year}{1996}).

\bibitem[{\citenamefont{Phillips}(2003)}]{Phillips_2003}
\bibinfo{author}{\bibfnamefont{J.~C.} \bibnamefont{Phillips}},
  \bibinfo{journal}{cond-mat/0304501}  (\bibinfo{year}{2003}),
  \bibinfo{note}{unpublished}.

\bibitem[{\citenamefont{Phillips}(2000)}]{Phillips_2000}
\bibinfo{author}{\bibfnamefont{J.~C.} \bibnamefont{Phillips}},
  \bibinfo{journal}{Physica C} \textbf{\bibinfo{volume}{340}},
  \bibinfo{pages}{292} (\bibinfo{year}{2000}).

\bibitem[{\citenamefont{Phillips and Vandenberg}(1997)}]{Phillips_97}
\bibinfo{author}{\bibfnamefont{J.~C.} \bibnamefont{Phillips}} \bibnamefont{and}
  \bibinfo{author}{\bibfnamefont{J.~M.} \bibnamefont{Vandenberg}},
  \bibinfo{journal}{J. Phys.: Condens. Matter} \textbf{\bibinfo{volume}{9}},
  \bibinfo{pages}{L251} (\bibinfo{year}{1997}).

\bibitem[{\citenamefont{Donati et~al.}(1999)\citenamefont{Donati, Glotzer,
  Poole, Kob, and Plimpton}}]{Donati_99}
\bibinfo{author}{\bibfnamefont{C.}~\bibnamefont{Donati}},
  \bibinfo{author}{\bibfnamefont{S.~C.} \bibnamefont{Glotzer}},
  \bibinfo{author}{\bibfnamefont{P.~H.} \bibnamefont{Poole}},
  \bibinfo{author}{\bibfnamefont{W.}~\bibnamefont{Kob}}, \bibnamefont{and}
  \bibinfo{author}{\bibfnamefont{S.~J.} \bibnamefont{Plimpton}},
  \bibinfo{journal}{Phys. Rev. E} \textbf{\bibinfo{volume}{60}},
  \bibinfo{pages}{3107} (\bibinfo{year}{1999}).

\bibitem[{\citenamefont{Debenedetti and Stillinger}(2001)}]{Debenedetti_01}
\bibinfo{author}{\bibfnamefont{P.~G.} \bibnamefont{Debenedetti}}
  \bibnamefont{and} \bibinfo{author}{\bibfnamefont{F.~H.}
  \bibnamefont{Stillinger}}, \bibinfo{journal}{Nature}
  \textbf{\bibinfo{volume}{410}}, \bibinfo{pages}{259} (\bibinfo{year}{2001}).

\bibitem[{\citenamefont{Garrahan and Chandler}(2002)}]{Garrahan_2002}
\bibinfo{author}{\bibfnamefont{J.~P.} \bibnamefont{Garrahan}} \bibnamefont{and}
  \bibinfo{author}{\bibfnamefont{D.}~\bibnamefont{Chandler}},
  \bibinfo{journal}{Phys. Rev. Lett.} \textbf{\bibinfo{volume}{89}},
  \bibinfo{pages}{035704} (\bibinfo{year}{2002}).

\bibitem[{\citenamefont{Mansfield}(2002)}]{Mansfield_2002}
\bibinfo{author}{\bibfnamefont{M.~L.} \bibnamefont{Mansfield}},
  \bibinfo{journal}{Phys. Rev. E} \textbf{\bibinfo{volume}{66}},
  \bibinfo{pages}{016101} (\bibinfo{year}{2002}).

\bibitem[{\citenamefont{S.~Bhattacharyya and
  Bagchi}(2002)}]{Bhattacharyya_2002}
\bibinfo{author}{\bibfnamefont{A.~M.} \bibnamefont{S.~Bhattacharyya}}
  \bibnamefont{and} \bibinfo{author}{\bibfnamefont{B.}~\bibnamefont{Bagchi}},
  \bibinfo{journal}{J. Chem. Phys.} \textbf{\bibinfo{volume}{117}},
  \bibinfo{pages}{2741} (\bibinfo{year}{2002}).

\bibitem[{\citenamefont{Hunt et~al.}(2002)\citenamefont{Hunt, Gade, and
  Mousseau}}]{Hunt_2002}
\bibinfo{author}{\bibfnamefont{E.}~\bibnamefont{Hunt}},
  \bibinfo{author}{\bibfnamefont{P.}~\bibnamefont{Gade}}, \bibnamefont{and}
  \bibinfo{author}{\bibfnamefont{N.}~\bibnamefont{Mousseau}},
  \bibinfo{journal}{Europhys. Lett.} \textbf{\bibinfo{volume}{60}},
  \bibinfo{pages}{827} (\bibinfo{year}{2002}).

\bibitem[{\citenamefont{Richert}(2002)}]{Richert_2002}
\bibinfo{author}{\bibfnamefont{R.}~\bibnamefont{Richert}}, \bibinfo{journal}{J.
  Phys.: Condens. Matter} \textbf{\bibinfo{volume}{14}}, \bibinfo{pages}{R703}
  (\bibinfo{year}{2002}).

\bibitem[{\citenamefont{Sillescu}(1999)}]{Sillescu_99}
\bibinfo{author}{\bibfnamefont{H.}~\bibnamefont{Sillescu}},
  \bibinfo{journal}{J. Non-Cryst. Solids} \textbf{\bibinfo{volume}{243}},
  \bibinfo{pages}{81} (\bibinfo{year}{1999}).

\bibitem[{\citenamefont{Cristanti and Ritort}(2002)}]{cristanti02}
\bibinfo{author}{\bibfnamefont{A.}~\bibnamefont{Cristanti}} \bibnamefont{and}
  \bibinfo{author}{\bibfnamefont{F.}~\bibnamefont{Ritort}},
  \bibinfo{journal}{Philos. Mag. B} \textbf{\bibinfo{volume}{82}},
  \bibinfo{pages}{143} (\bibinfo{year}{2002}).

\bibitem[{\citenamefont{Chong et~al.}(2001)\citenamefont{Chong, G\"otze, and
  Mayr}}]{chong01}
\bibinfo{author}{\bibfnamefont{S.-H.} \bibnamefont{Chong}},
  \bibinfo{author}{\bibfnamefont{W.}~\bibnamefont{G\"otze}}, \bibnamefont{and}
  \bibinfo{author}{\bibfnamefont{M.~R.} \bibnamefont{Mayr}},
  \bibinfo{journal}{Phys. Rev. E} \textbf{\bibinfo{volume}{64}},
  \bibinfo{pages}{011503} (\bibinfo{year}{2001}).

\bibitem[{\citenamefont{Chong and G\"otze}(2002)}]{chong02}
\bibinfo{author}{\bibfnamefont{S.-H.} \bibnamefont{Chong}} \bibnamefont{and}
  \bibinfo{author}{\bibfnamefont{W.}~\bibnamefont{G\"otze}},
  \bibinfo{journal}{Phys. Rev. E} \textbf{\bibinfo{volume}{65}},
  \bibinfo{pages}{051201} (\bibinfo{year}{2002}).

\bibitem[{\citenamefont{Vollmay-Lee et~al.}(2002)\citenamefont{Vollmay-Lee,
  Kob, Binder, and Zippelius}}]{VollmayrLee_2002}
\bibinfo{author}{\bibfnamefont{K.}~\bibnamefont{Vollmay-Lee}},
  \bibinfo{author}{\bibfnamefont{W.}~\bibnamefont{Kob}},
  \bibinfo{author}{\bibfnamefont{K.}~\bibnamefont{Binder}}, \bibnamefont{and}
  \bibinfo{author}{\bibfnamefont{A.}~\bibnamefont{Zippelius}},
  \bibinfo{journal}{J. Chem. Phys.} \textbf{\bibinfo{volume}{116}},
  \bibinfo{pages}{5158} (\bibinfo{year}{2002}).

\bibitem[{\citenamefont{Caprion et~al.}(2000)\citenamefont{Caprion, Matsui, and
  Schober}}]{Caprion_2000}
\bibinfo{author}{\bibfnamefont{D.}~\bibnamefont{Caprion}},
  \bibinfo{author}{\bibfnamefont{J.}~\bibnamefont{Matsui}}, \bibnamefont{and}
  \bibinfo{author}{\bibfnamefont{H.~R.} \bibnamefont{Schober}},
  \bibinfo{journal}{Phys. Rev. Lett.} \textbf{\bibinfo{volume}{85}},
  \bibinfo{pages}{4293} (\bibinfo{year}{2000}).

\bibitem[{\citenamefont{Allegrini et~al.}(1999)\citenamefont{Allegrini,
  Douglas, and Glotzer}}]{Allegrini_1999}
\bibinfo{author}{\bibfnamefont{P.}~\bibnamefont{Allegrini}},
  \bibinfo{author}{\bibfnamefont{J.~F.} \bibnamefont{Douglas}},
  \bibnamefont{and} \bibinfo{author}{\bibfnamefont{C.}~\bibnamefont{Glotzer}},
  \bibinfo{journal}{Phys. Rev. E} \textbf{\bibinfo{volume}{60}},
  \bibinfo{pages}{5714} (\bibinfo{year}{1999}).

\bibitem[{\citenamefont{Doliwa and Heuer}(2003)}]{Doliwa_2003}
\bibinfo{author}{\bibfnamefont{B.}~\bibnamefont{Doliwa}} \bibnamefont{and}
  \bibinfo{author}{\bibfnamefont{A.}~\bibnamefont{Heuer}},
  \bibinfo{journal}{Phys. Rev. E} \textbf{\bibinfo{volume}{67}},
  \bibinfo{pages}{030501} (\bibinfo{year}{2003}).

\bibitem[{\citenamefont{Denny et~al.}(2003)\citenamefont{Denny, Reichman, and
  Bouchaud}}]{Denny_2003}
\bibinfo{author}{\bibfnamefont{R.~A.} \bibnamefont{Denny}},
  \bibinfo{author}{\bibfnamefont{D.~R.} \bibnamefont{Reichman}},
  \bibnamefont{and} \bibinfo{author}{\bibfnamefont{J.~P.}
  \bibnamefont{Bouchaud}}, \bibinfo{journal}{Phys. Rev. Let..}
  \textbf{\bibinfo{volume}{90}}, \bibinfo{pages}{025503}
  (\bibinfo{year}{2003}).

\bibitem[{\citenamefont{Xia and Wolynes}(2001)}]{Xia_2001}
\bibinfo{author}{\bibfnamefont{X.~Y.} \bibnamefont{Xia}} \bibnamefont{and}
  \bibinfo{author}{\bibfnamefont{P.~G.} \bibnamefont{Wolynes}},
  \bibinfo{journal}{Phys. Rev. Lett.} \textbf{\bibinfo{volume}{86}},
  \bibinfo{pages}{5526} (\bibinfo{year}{2001}).

\bibitem[{\citenamefont{Simdyankin and Mousseau}(2002)}]{Simdyankin_2002_PRE}
\bibinfo{author}{\bibfnamefont{S.}~\bibnamefont{Simdyankin}} \bibnamefont{and}
  \bibinfo{author}{\bibfnamefont{N.}~\bibnamefont{Mousseau}},
  \bibinfo{journal}{Phys. Rev. E} \textbf{\bibinfo{volume}{66}},
  \bibinfo{pages}{066205} (\bibinfo{year}{2002}), \eprint{condmat/0208487}.

\bibitem[{\citenamefont{Godr\`{e}che and Luck}(2001)}]{Godreche_2001}
\bibinfo{author}{\bibfnamefont{C.}~\bibnamefont{Godr\`{e}che}}
  \bibnamefont{and} \bibinfo{author}{\bibfnamefont{J.~M.} \bibnamefont{Luck}},
  \bibinfo{journal}{J. Stat. Phys.} \textbf{\bibinfo{volume}{104}},
  \bibinfo{pages}{489} (\bibinfo{year}{2001}).

\bibitem[{\citenamefont{Hughes}(1995)}]{Hughes_RW}
\bibinfo{author}{\bibfnamefont{B.~D.} \bibnamefont{Hughes}},
  \emph{\bibinfo{title}{Random Walks and Random Environments}},
  vol.~\bibinfo{volume}{1} (\bibinfo{publisher}{Clarendon Press},
  \bibinfo{address}{Oxford}, \bibinfo{year}{1995}).

\bibitem[{\citenamefont{Barkema et~al.}(2001)\citenamefont{Barkema, Biswas, and
  van Beijeren}}]{Barkema_2001}
\bibinfo{author}{\bibfnamefont{G.}~\bibnamefont{Barkema}},
  \bibinfo{author}{\bibfnamefont{P.}~\bibnamefont{Biswas}}, \bibnamefont{and}
  \bibinfo{author}{\bibfnamefont{H.}~\bibnamefont{van Beijeren}},
  \bibinfo{journal}{Phys. Rev. Lett.} \textbf{\bibinfo{volume}{87}},
  \bibinfo{pages}{170601} (\bibinfo{year}{2001}).

\bibitem[{\citenamefont{J.~N.~Roux}(1989)}]{Roux_1989}
\bibinfo{author}{\bibnamefont{J.~N.~Roux}},
\bibinfo{author}{\bibfnamefont{J.~L.~Barrat}}, 
  \bibnamefont{and}  
\bibinfo{author}{\bibfnamefont{J.-P.~Hansen}}, 
\bibinfo{journal}{J. Phys.: Condens. Matter}
  \textbf{\bibinfo{volume}{1}}, \bibinfo{pages}{7171} (\bibinfo{year}{1989}).

\bibitem[{\citenamefont{G\"otze and Sj\"ogren}(1992)}]{Gotze_92}
\bibinfo{author}{\bibfnamefont{W.}~\bibnamefont{G\"otze}} \bibnamefont{and}
  \bibinfo{author}{\bibfnamefont{L.}~\bibnamefont{Sj\"ogren}},
  \bibinfo{journal}{Rep. Prog. Phys.} \textbf{\bibinfo{volume}{55}},
  \bibinfo{pages}{241} (\bibinfo{year}{1992}).

\bibitem[{\citenamefont{Ogielski}(1985)}]{Ogielski_1985}
\bibinfo{author}{\bibfnamefont{A.~T.} \bibnamefont{Ogielski}},
  \bibinfo{journal}{Phys. Rev. B} \textbf{\bibinfo{volume}{32}},
  \bibinfo{pages}{7384} (\bibinfo{year}{1985}).

\bibitem[{\citenamefont{Campbell}(1985)}]{Campbell_85}
\bibinfo{author}{\bibfnamefont{I.}~\bibnamefont{Campbell}},
  \bibinfo{journal}{J. Physique Lett.} \textbf{\bibinfo{volume}{46}},
  \bibinfo{pages}{L1159} (\bibinfo{year}{1985}).

\bibitem[{\citenamefont{de~Almeida et~al.}(2000)\citenamefont{de~Almeida,
  Lemke, and Campbell}}]{Almeida_2000}
\bibinfo{author}{\bibfnamefont{R.}~\bibnamefont{de~Almeida}},
  \bibinfo{author}{\bibfnamefont{N.}~\bibnamefont{Lemke}}, \bibnamefont{and}
  \bibinfo{author}{\bibfnamefont{I.}~\bibnamefont{Campbell}},
  \bibinfo{journal}{Eur. Phys. J. B} \textbf{\bibinfo{volume}{18}},
  \bibinfo{pages}{513} (\bibinfo{year}{2000}).

\bibitem[{\citenamefont{Jund et~al.}(2001)\citenamefont{Jund, Jullien, and
  Campbell}}]{Jund_2001}
\bibinfo{author}{\bibfnamefont{P.}~\bibnamefont{Jund}},
  \bibinfo{author}{\bibfnamefont{R.}~\bibnamefont{Jullien}}, \bibnamefont{and}
  \bibinfo{author}{\bibfnamefont{I.}~\bibnamefont{Campbell}},
  \bibinfo{journal}{Phys. Rev. E} \textbf{\bibinfo{volume}{63}},
  \bibinfo{pages}{036131} (\bibinfo{year}{2001}).

\bibitem[{\citenamefont{Ott}(1993)}]{Ott_CDS}
\bibinfo{author}{\bibfnamefont{E.}~\bibnamefont{Ott}},
  \emph{\bibinfo{title}{Chaos in Dynamical Systems}}
  (\bibinfo{publisher}{Cambridge University Press},
  \bibinfo{address}{Cambridge}, \bibinfo{year}{1993}).

\bibitem[{\citenamefont{Feller}(1970)}]{Feller_IPTA}
\bibinfo{author}{\bibfnamefont{W.}~\bibnamefont{Feller}},
  \emph{\bibinfo{title}{An Introduction to Probability Theory and its
  Applications}}, vol.~\bibinfo{volume}{1} (\bibinfo{publisher}{Wiley},
  \bibinfo{address}{New York}, \bibinfo{year}{1970}), \bibinfo{edition}{3rd}
  ed., \bibinfo{note}{revised Printing}.

\bibitem[{\citenamefont{Gielis and Maes}(1995)}]{Gielis_95}
\bibinfo{author}{\bibfnamefont{G.}~\bibnamefont{Gielis}} \bibnamefont{and}
  \bibinfo{author}{\bibfnamefont{C.}~\bibnamefont{Maes}},
  \bibinfo{journal}{Europhys. Lett.} \textbf{\bibinfo{volume}{31}},
  \bibinfo{pages}{1} (\bibinfo{year}{1995}).

\bibitem[{\citenamefont{Godr\`{e}che and Luck}(2001)}]{godreche2002}
\bibinfo{author}{\bibfnamefont{C.}~\bibnamefont{Godr\`{e}che}}
  \bibnamefont{and} \bibinfo{author}{\bibfnamefont{J.~M.} \bibnamefont{Luck}},
  \bibinfo{journal}{private communication}.

\bibitem{footnote1} The following analysis also holds if $\rho(t)$ is a L\'evy
  law.

\bibitem[{\citenamefont{Devroye}(1986)}]{Devroye_NURVG}
\bibinfo{author}{\bibfnamefont{L.}~\bibnamefont{Devroye}},
  \emph{\bibinfo{title}{Non-uniform random variate generation}}
  (\bibinfo{publisher}{Springer-Verlag}, \bibinfo{address}{New York},
  \bibinfo{year}{1986}).

\bibitem[{\citenamefont{DiDonato and A.H.~Morris}(1987)}]{DiDonato_87}
\bibinfo{author}{\bibfnamefont{A.}~\bibnamefont{DiDonato}} \bibnamefont{and}
  \bibinfo{author}{\bibfnamefont{J.}~\bibnamefont{A.H.~Morris}},
  \bibinfo{journal}{ACM Trans. Math. Softw.} \textbf{\bibinfo{volume}{13}},
  \bibinfo{pages}{318} (\bibinfo{year}{1987}),
  \urlprefix\url{http://doi.acm.org/10.1145/29380.214348}.

\bibitem[{\citenamefont{DiDonato and A.H.~Morris}(1986)}]{DiDonato_86}
\bibinfo{author}{\bibfnamefont{A.}~\bibnamefont{DiDonato}} \bibnamefont{and}
  \bibinfo{author}{\bibfnamefont{J.}~\bibnamefont{A.H.~Morris}},
  \bibinfo{journal}{ACM Trans. Math. Softw.} \textbf{\bibinfo{volume}{12}},
  \bibinfo{pages}{377} (\bibinfo{year}{1986}).

\bibitem[{\citenamefont{Doye et~al.}(2003)\citenamefont{Doye, Wales,
  Zetterling, and Dzugutov}}]{Doye_2002_Z1and2}
\bibinfo{author}{\bibfnamefont{J.~P.~K.} \bibnamefont{Doye}},
  \bibinfo{author}{\bibfnamefont{D.~J.} \bibnamefont{Wales}},
  \bibinfo{author}{\bibfnamefont{F.~H.~M.} \bibnamefont{Zetterling}},
  \bibnamefont{and} \bibinfo{author}{\bibfnamefont{M.}~\bibnamefont{Dzugutov}},
  \bibinfo{journal}{J. Chem. Phys.} \textbf{\bibinfo{volume}{118}},
  \bibinfo{pages}{2792} (\bibinfo{year}{2003}), \eprint{cond-mat/0205374}.

\bibitem[{\citenamefont{Dzugutov}(1992)}]{Dzugutov_92}
\bibinfo{author}{\bibfnamefont{M.}~\bibnamefont{Dzugutov}},
  \bibinfo{journal}{Phys. Rev. A} \textbf{\bibinfo{volume}{46}},
  \bibinfo{pages}{R2984} (\bibinfo{year}{1992}).

\bibitem[{\citenamefont{Dzugutov et~al.}(2002)\citenamefont{Dzugutov,
  Simdyankin, and Zetterling}}]{Dzugutov_2001}
\bibinfo{author}{\bibfnamefont{M.}~\bibnamefont{Dzugutov}},
  \bibinfo{author}{\bibfnamefont{S.~I.} \bibnamefont{Simdyankin}},
  \bibnamefont{and} \bibinfo{author}{\bibfnamefont{F.~H.~M.}
  \bibnamefont{Zetterling}}, \bibinfo{journal}{Phys. Rev. Lett.}
  \textbf{\bibinfo{volume}{89}}, \bibinfo{pages}{195701}
  (\bibinfo{year}{2002}).

\bibitem{footnote2} Following Refs.~\cite{Dzugutov_92,Doye_2002_Z1and2}, we use
reduced Lennard-Jones units \cite{Allen_CSL} for all quantities here.

\bibitem[{\citenamefont{Hansen and McDonald}(1986)}]{Hansen_TSL}
\bibinfo{author}{\bibfnamefont{J.-P.} \bibnamefont{Hansen}} \bibnamefont{and}
  \bibinfo{author}{\bibfnamefont{I.~R.} \bibnamefont{McDonald}},
  \emph{\bibinfo{title}{Theory of Simple Liquids}}
  (\bibinfo{publisher}{Academic Press}, \bibinfo{address}{London},
  \bibinfo{year}{1986}), \bibinfo{edition}{2nd} ed.

\bibitem[{\citenamefont{Dzugutov}(1994)}]{Dzugutov_94}
\bibinfo{author}{\bibfnamefont{M.}~\bibnamefont{Dzugutov}},
  \bibinfo{journal}{Europhys. Lett.} \textbf{\bibinfo{volume}{26}},
  \bibinfo{pages}{533} (\bibinfo{year}{1994}).

\bibitem[{\citenamefont{Rabani et~al.}(1999)\citenamefont{Rabani, Gezelter, and
  Berne}}]{Rabani_99}
\bibinfo{author}{\bibfnamefont{E.}~\bibnamefont{Rabani}},
  \bibinfo{author}{\bibfnamefont{J.~D.} \bibnamefont{Gezelter}},
  \bibnamefont{and} \bibinfo{author}{\bibfnamefont{B.}~\bibnamefont{Berne}},
  \bibinfo{journal}{Phys. Rev. Lett.} \textbf{\bibinfo{volume}{82}},
  \bibinfo{pages}{3649} (\bibinfo{year}{1999}).

\bibitem[{\citenamefont{Kayser and Habbard}(1983)}]{Kayser_1983}
\bibinfo{author}{\bibfnamefont{R.~F.} \bibnamefont{Kayser}} \bibnamefont{and}
  \bibinfo{author}{\bibfnamefont{J.~B.} \bibnamefont{Habbard}},
  \bibinfo{journal}{Phys. Rev. Lett.} \textbf{\bibinfo{volume}{51}},
  \bibinfo{pages}{79} (\bibinfo{year}{1983}).

\bibitem[{\citenamefont{Grassberger and Procaccia}(1982)}]{Grassberger_82}
\bibinfo{author}{\bibfnamefont{P.}~\bibnamefont{Grassberger}} \bibnamefont{and}
  \bibinfo{author}{\bibfnamefont{I.}~\bibnamefont{Procaccia}},
  \bibinfo{journal}{J. Chem. Phys.} \textbf{\bibinfo{volume}{77}},
  \bibinfo{pages}{6281} (\bibinfo{year}{1982}).

\bibitem[{\citenamefont{Allen and Tildesley}(1987)}]{Allen_CSL}
\bibinfo{author}{\bibfnamefont{M.~P.} \bibnamefont{Allen}} \bibnamefont{and}
  \bibinfo{author}{\bibfnamefont{D.~J.} \bibnamefont{Tildesley}},
  \emph{\bibinfo{title}{Computer Simulation of Liquids}}
  (\bibinfo{publisher}{Clarendon Press}, \bibinfo{address}{Oxford},
  \bibinfo{year}{1987}).

\end{thebibliography}

\end{document}